\documentclass[a4paper,notitlepage]{article}
\usepackage{amsmath}
\usepackage{graphicx}
\begin{document}

\def\ft{\footnote}
\def\half{\frac{1}{2}}

\renewcommand{\thefootnote}{\arabic{footnote}}

\newcommand{\DE}{\ensuremath{\Delta_l^\mathrm{\!TE}}}
\newcommand{\DM}{\ensuremath{\Delta_l^\mathrm{\!TM}}}

\begin{titlepage}

\title{Finite temperature Casimir effect for a dilute ball
    satisfying $\epsilon\mu=1$ \thanks{Revised Version}} 

\author{%
I.~Brevik\ft{E-mail address:iver.h.brevik@mtf.ntnu.no}~~ and~%
T.~A.~Yousef\ft{E mail address: tarek.yousef@mtf.ntnu.no}\\
    \\
    Division of Applied Mechanics\\
    Norwegian University of Science and Technology\\
    N-7491 Trondheim, Norway.  }

\end{titlepage}

\maketitle

\begin{abstract}
The finite temperature Casimir free energy is calculated for a dielectric
ball of radius $a$  embedded in an infinite medium.  The condition
$\epsilon\mu=1$ is assumed for the inside/outside regions. Both the
Green function method and the mode summation method are considered,
and found to be equivalent. For a dilute medium we find, assuming a simple
"square" dispersion relation with an abrupt cutoff at imaginary frequency $\hat \omega= \omega_0$, 
the
high temperature Casimir free energy to be negative and proportional to
$x_0 \equiv \omega_0 a$. Also, a physically more realistic dispersion relation
involving spatial dispersion is considered, and is shown to lead to comparable results.  
\end{abstract}

\section{Introduction}

The Casimir effect has in recent years attracted considerable
interest. For reviews on this topic see, for example, Plunien
\emph{et al.}\cite{Plunien86}, Mostepanenko and
Trunov\cite{Mostepanenko97}, and Milton \cite{Milton99}. The problem
has been rather difficult to solve, once curved boundaries are present,
because of divergences in the formalism and insufficient contact with
experiments. In this paper we will consider a compact ball of radius a,
permittivity $\epsilon_1$ and permeability $\mu_1$, situated in an
infinite medium whose corresponding material constants are
$\epsilon_2$ and $\mu_2$.

We shall take 
\begin{equation}\label{eq:1}
\epsilon_1\mu_1 = \epsilon_2\mu_2 = 1
\end{equation}
\label{1}
so that the photon velocity is everywhere equal to $c$.

The relativistic condition (1) was, in the context of the Casimir
effect, to our knowledge first introduced by Brevik and Kolbenstvedt
\cite{Brevik82,Brevik84}. The reason why this particular condition was
imposed was that when calculating the Casimir surface force density on
the sphere, in this paper called $f$, one avoids thereby the
subtraction of a contact term: the contact term turns out to be zero
just as in the case of a singular spherical conducting sphere in a
vacuum \cite{Milton78}. Later, this kind of medium has been considered
from various viewpoints [7 - 19]. A common feature of most of these
papers is the use of macroscopic electrodynamics of continuous media.
 
The main purpose of the present paper is to consider the Casimir
effect for a compact ball satisfying the condition (1), at finite
temperatures. Finite temperature Casimir theory has been considered
previously by several authors, e.g., Lifshitz \cite{lifshitz55},
 Fierz \cite{fierz60}, Sauer \cite{sauer62}, Mehra \cite{mehra67},
 Brown and Maclay \cite{Brown69},
Milton {\it et al.} \cite {Milton78a}, Revzen {\it et al.} \cite
{Revzen97}, and Klich \cite{Klich00}.  Also the very recent extensive
treatise of Feinberg {\it et al.} \cite{Feinberg99} ought to be
mentioned, since it describes in detail the characteristic properties
of the high temperature limit, and contains also an extensive
reference list.

We start in the next section by considering, as a first step, the
$T=0$ theory, and demonstrate the equivalence between the Green
function method and the mode summation method. The equivalence has of
course to hold if the theory is to make sense physically, but the
equivalence is not so easy to see by a mere inspection of the
mathematics. A detailed calculation is desirable. In section 3 we generalize
 the Casimir energy expression obtained in \cite{Brevik98} to the fourth order 
in the diluteness parameter $\xi$; cf. Eq. (15) and the definition in Eq. (7). In section 4  we consider the
 Casimir free energy $F(T)$ at finite $T$,
emphasizing the difference between $F(T)$ and the Casimir energy $E(T)$, and
give an order of magnitude estimate of $F(T)$ in the case of a dilute
medium endowed with a "square" dispersion relation implying a sharp
high frequence cutoff at imaginary frequency $\hat \omega=\omega_0$.  The presence of a
dispersion-induced strong, negative, cutoff-dependent contribution to
the Casimir energy (or free energy, for finite $T$), seems first to
have been pointed out by Candelas \cite{Candelas82}, for the case of a
perfectly conducting shell.

At high temperatures we find $F(T)$ to be negative, decreasing linearly
with $T$. This agrees with results found previously in \cite{Revzen97}
and \cite{Feinberg99}. The Casimir energy $E(T)$ itself tends to zero at
high temperatures.

\section{Equivalence between the Green function approach and the mode
  sum approach when $T=0$.  Nondispersive medium}
\label{sec:2}

In this section we consider first the Casimir surface force density
$f$ at zero temperature, for a nondispersive medium.  There are in principle
 two different methods for calculating $f$ (more precisely, the part of $f$ that is
independent of the high frequency cutoff):

The first method is to make use of the Green functions. There are two
scalar Green functions, corresponding to two different electromagnetic
modes. Again, this method can be carried out in one of two different
ways: 
\begin{enumerate}
\item First, one can evaluate $f$ as the difference between the radial
  Maxwell stress tensor components on the outside and inside of the
  surface $r=a$. We have to subtract off the volume terms in the
  scalar Green functions, so that only the boundary dependent terms
  are left. This regularization ensures that the surface force goes to
  zero when $a\to\infty$, which is as we expect on physical
  grounds. Also, this regularization permits us to calculate the
  Casimir  energy $E$ once $f$ is known, by the formula 
  $\partial E/\partial a = -4\pi a^2 f $. (At $T=0$, $E$ is the same as the free energy $F$.)
\item Alternatively, one can calculate the energy $E$ by integrating
  the Maxwell energy density over the volume, retaining, as above,
  only the boundary dependent terms in the Green functions. The
  outcome of these two approaches are of course in agreement.
\end{enumerate}

The Green function method was first employed by Milton, for the case
of an ordinary dielectric ball \cite{Milton80}, and was later applied
to the case of an $\epsilon\mu=1$ ball in Ref. \cite{Brevik82}.

The result of the calculation can be given as follows. After a complex
frequency rotation $\omega \to i \hat\omega$ we can express all
physical quantities as integrals over frequencies $ \hat \omega$ along the
imaginary frequency axis.\footnote{ This frequency rotation is a 
bit more involved than it might seem at first sight, since the Hankel functions
imply that there are singularities in the lower half of the complex frequency plane. Physically,
 these singularities occur because we assume an infinite exterior region (thus, no large exterior
 sphere making all eigenfrequencies real). The legitimacy of the complex frequency rotation in the presence of these complexities follows from a more careful derivation; cf. the recent discussion in Ref. \cite{brevik00}.} 
The cutoff dependence is omitted (it
disappears in the regularization procedure). We define Riccati-Bessel
functions $s_l$ and $e_l$ by ($x= \hat k a )$ 
\begin{equation} \label{eq:2}
s_l(x) = \sqrt\frac{\pi x}2 I_\nu(x), \qquad
e_l(x) = \sqrt\frac{2 x}\pi K_\nu(x),
\end{equation}
\label{2}
corresponding to the Wronskian $W \{ s_l, e_l  \}= -1$. 
Here $\nu=l+\half$; $I_\nu$ and $K_\nu$ are modified Bessel
functions. We introduce two dispersion functions \DE  and \DM
corresponding to the TE and TM modes:
\begin{eqnarray} \label{eq:3and4}
\DE(x)=\sqrt{\epsilon_1\mu_2}s_l'(x)e_l(x)-\sqrt{\epsilon_2\mu_1}s_l(x)e_l'(x),\\
\DM(x)=\sqrt{\epsilon_2\mu_1}s_l'(x)e_l(x)-\sqrt{\epsilon_1\mu_2}s_l(x)e_l'(x).
\end{eqnarray}
\label{4}

The Casimir energy as derived in \cite{Brevik82} on the basis of the
Green function can be written as 
\begin{equation} \label{eq:5}
E= \frac 1{2a}\frac{(\mu_1-\mu_2)^2}{\mu_1\mu_2}
\int_{-\infty}^\infty\frac{dy}{2\pi}\, e^{i\delta y}
\sum_{l=1}^\infty\frac{2l+1}4 \frac x{\DE(x)\DM(x)}
\left[\lambda_l^2(x)\right]'.
\end{equation}
\label{5}
Here $y=\hat\omega a$, $x=|y|$; the time difference $\tau=t-t'$ has
been Euclidean rotated, $\tau\to i\hat\tau$, and $\delta=\hat\tau/a$
is the time-splitting parameter. Moreover, we have defined the
function $\lambda_l(x)$ as $\lambda_l(x)=[s_l(x)e_l(x)]'$   [In Ref
\cite{Brevik82} we used the symbols $D_l(x)$ and $\tilde D_l(x)$; the
relations between these quantities and those defined by
(3) and (4) are $D_l(x)=-\sqrt{\mu_1\mu_2}\DE(x)$ and $\tilde
D_l(x)=-\sqrt{\mu_1\mu_2}\DM(x)$. ]

Consider next the second method, which consists in summing over the
modes by means of contour integration. This method was recently used
in Ref. \cite{Brevik98}, and is in turn based on the technique
developed by Lambiase and Nesterenko \cite{Lambiase96} and Nesterenko
and Pirozhenko \cite{Nesterenko97a}. One gets (when we supply the same
parameter $\delta$ as above) 
\begin{equation} \label{eq:6}
E= -\frac 1{2a}\sum_{l=1}^\infty (2l+1)
\int_{-\infty}^\infty\frac{dy}{2\pi}\, e^{i\delta y}
  x \frac d{dx} \ln \left[\DE(x)\DM(x)\right].
\end{equation}
\label{6}
Introducing the parameter $\xi$ by
\begin{equation} \label{eq:7}
\xi = \frac{\epsilon_1-\epsilon_2}{\epsilon_1+\epsilon_2}
= \frac{\mu_2-\mu_1}{\mu_2+\mu_1},
\end{equation}
\label{7}
the expression (6) can alternatively be written as 
\begin{equation} \label{eq:8}
E= -\frac 1{2a}\sum_{l=1}^\infty (2l+1)
\int_{-\infty}^\infty\frac{dy}{2\pi}\, e^{i\delta y}
  x \frac d{dx} \ln \left[1-\xi^2\lambda_l^2(x)\right].
\end{equation}
\label{8}

The two methods thus lead to two expressions for the Casimir energy,
given by Eqs. (\ref{eq:5})  and (\ref{eq:6}). The expressions, of
course, have to be equal, but to see the equality is not quite trivial
by mere inspection. The equality is tantamount to the equality
\begin{equation} \label{eq:9}
\left[\DE(x)\DM(x)\right]'= 
-\frac{(\mu_1-\mu_2)^2}{4\mu_1\mu_2} \left[\lambda_l^2(x)\right]'.
\end{equation}
\label{9}

Now it turns out that this equality is not so difficult to prove after
all if we insert the expressions (3) and (4) on the left hand side and
take into account the above mentioned Wronskian between $s_l$ and
$e_l$. Actually, the expression (6) was also made use of in
the Green-function calculation of Milton and Ng \cite{Milton97}.

\section{A remark on the dilute ball when T=0} 

Before leaving the zero temperature theory let us make a brief remark
on the nondispersive dilute ball, meaning that 
\begin{equation} \label{eq:10} |\xi| \ll 1 . \end{equation}
\label{10}

The most practical expression to make use of for the Casimir energy,
is that of Eq.(\ref{eq:8}). We expand the logarithm to fourth order
in $\xi$:
\begin{equation} \label{eq:11}
\ln(1-\xi^2\lambda_l^2)= -\xi^2\lambda^2_l(x)-\half\xi^4\lambda_l^4+{\cal O}(\xi^6),
\end{equation}
\label{11}
and make the following steps:
\begin{enumerate}
\item put $\delta=0$;
\item perform a partial integration in Eq.(8);
\item interchange summation and integration signs.
\end{enumerate}
Then, omitting the $\mathcal{ O} (\xi^6)$ terms in (\ref{eq:11}),
\begin{equation} 
E= -\frac{\xi^2}{\pi a} \sum_{l=1}^\infty\nu\int_0^\infty dx\,\lambda_l^2(x)
   -\frac{\xi^4}{2\pi a}\sum_{l=1}^\infty\nu\int_0^\infty dx\,\lambda_l^4(x),
\end{equation}
\label{12}
which can be processed further using the uniform asymptotic expansion
for $s_l$ and $e_l$ up to $\mathcal {O} (1/\nu^4)$, in the same way as in
Refs.~\cite{Brevik98,Brevik87}. There is one single divergent term
in Eq. (12), which can be regularized 
by means of the Riemann zeta function, $\zeta(s)$. The only formula needed in practice is
\begin{equation} 
\sum_{l=1}^\infty\nu^{-s}=\left(2^s-1\right)\zeta(s)-2^s.
\end{equation}
\label{13}

Moreover, we use the summations $\sum_1^\infty\nu^{-2}=\half\pi^2 -4$,~~
$\sum_1^\infty\nu^{-4}=\frac 1 6 \pi^4 -16$, as well as the integral
\begin{equation} \label{eq:14}
 \int_0^\infty (1 +z^2)^{-p / 2} \, dz =
 \half \,  \mathrm B \left( \half, \frac {p-1} 2 \right)
\end{equation}
\label{14}
($\mathrm B$ is the beta function), to get 
\begin{equation} 
\begin{split}
E=&\frac{3 \xi^2}{64 a}
\left[1+\frac 9{128}\left(\half\pi^2-4\right)-\frac{423}{16384}\left(\frac 1 6
  \pi^4-16\right)\right]\\
&-\frac{63 \xi^4}{16384 a}
\left[\left(\half\pi^2-4\right)-\frac{351}{1792}\left(\frac {1}{ 6}\pi^4 -16 \right)\right].
\end{split}
\end{equation}
\label{15}

This expression generalizes that of equation (3.10) in \cite{Brevik98} up to the
fourth order in $\xi$. Numerically, the expressions between square
parentheses in (15) are respectively 1.05966 and 0.88880. The
fourth order contribution to the Casimir energy is thus weak and
negative, corresponding to an attractive force component.

In connection with this calculation, we wish to emphasise two points.
First, we may avoid the time-splitting  parameter $\delta$ from the
beginning, using zeta function regularization (here the Riemann
function) instead. Secondly, the convergence parameter $s$ that was
for mathematical reasons introduced in \cite{Brevik98}, can simply be
omitted. At least from a pragmatic point of view, the nondispersive
medium becomes quite simple.

\section{Dispersive medium: Finite temperatures} \label{sec:4}

We now take into account dispersive properties of the two media,
meaning that $\epsilon=\epsilon (\omega)$, $\mu = \mu (\omega)$. We shall still assume that $\epsilon (\omega) \mu(\omega)=1$, so that the velocity of photons in either medium is equal to the velocity of light. After frequency rotation, along the imaginary frequency axis, $\epsilon = \epsilon (i\hat \omega)$, $\mu = \mu (i\hat \omega)$.

Some care ought to be taken when considering the dispersive formalism, since frequency derivatives of the material constants are no longer ignorable straightaway. Let us first consider the $T=0$ total zero-point energy, which we shall call $E_{I+II}$, of the inner (I) and the outer (II) regions. This energy is simply given by Eq. (6), which we repeat here with a slightly generalized notation:
\begin{equation} 
E= -\frac 1{2a}\sum_{l=1}^\infty (2l+1)
\int_{-\infty}^\infty\frac{dy}{2\pi}\, e^{i\delta y}
  x \frac d{dx} \ln \left[\DE(i\hat \omega, a)\DM(i\hat \omega,a)\right].
\end{equation}
\label{16}
The reason for this equality is that the two dispersion functions $\Delta_l^{TE}$ and $\Delta_l^{TM}$, as given by the dispersive generalization of Eqs. (3) and (4), retain their physical meaning also in the presence of dispersion. Thus the TE and TM modes are still determined by $\Delta_l^{TE}=0$ and $\Delta_l^{TM}=0$  respectively. The expression (16) needs to be regularized: we subtract off the zero-point energy $E_{uniform}$ corresponding to the limit $a \rightarrow \infty$;  cf. \cite{Brevik98}. As $s_l(x)=\frac{1}{2}e^x$ and $e_l(x)=e^{-x}$ for large $x$, we get from (3) and (4)
\begin{equation}
\Delta_l^{TE}(i\hat \omega, a\rightarrow \infty)=\Delta_l^{TM}(i\hat \omega, a\rightarrow\infty)
=\frac{1}{2}(\sqrt{\epsilon_1\mu_2}+\sqrt{\epsilon_2\mu_1}).
\end{equation}
\label{17}
The Casimir energy $E=E_{I+II}-E_{uniform}$ becomes accordingly, at $T=0$,
\begin{equation} 
E= -\frac 1{2a}\sum_{l=1}^\infty (2l+1)
\int_{-\infty}^\infty\frac{dy}{2\pi}\, e^{i\delta y}
  x \frac d{dx} \ln \left[ \frac{4 \DE(i\hat \omega, a)\DM(i\hat \omega,a)}
 {(\sqrt{\epsilon_1\mu_2}+\sqrt{\epsilon_2\mu_1})^2} \right].
\end{equation}
\label{18}
Observing that $1+\lambda_l=2s_l' e_l$,  $1-\lambda_l=-2s_le_l'$ we find (when reverting to the notation $\Delta_l^{TE,TM}(x)$ instead of $\Delta_l^{TE,TM}(i\hat\omega,a)$) that
\begin{equation}
\Delta_l^{TE}(x)\Delta_l^{TM}(x)=\frac{1}{4}(\sqrt{\epsilon_1\mu_2}+\sqrt{\epsilon_2\mu_1})^2
[1-\xi^2\lambda_l^2(x)],
\end{equation}
\label{19}
which finally means that, at $T=0$,
\begin{equation}
E= -\frac 1{2a}\sum_{l=1}^\infty (2l+1)
\int_{-\infty}^\infty\frac{dy}{2\pi}\, 
  x \frac d{dx} \ln \left[ 1-\xi^2(x)\lambda_l^2(x) \right].
\end{equation}
\label{20}
It is rather remarkable that this simple formula continues to hold even in the presence of arbitrary frequency dispersion. Since dispersion implies that we introduce a physically based high frequency cutoff, we expect that there is no longer any need for a time splitting parameter. We have therefore put $\delta=0$ in the expression (20). We will henceforth restrict ourselves to the first ( i. e., the second order) term in the expansion (11).

In this section we adopt the simplest imaginable dispersion relation 
(a "square" form): we let the inner medium (a ball) correspond to
\begin{equation}
\mu(i\hat\omega)= 
\begin{cases} 
\mu= \text{const},         \qquad& \hat\omega\le\omega_0 \\
1,                     \qquad&\hat\omega > \omega_0,
\end{cases}
\end{equation}
\label{21}
whereas the outer medium is a vacuum. It is to be noted here
that the general thermodynamic law saying that the susceptibility has to
decrease monotonically along the positive imaginary frequency axis
\cite{Landau84}, is satisfied by this curve for $\epsilon (i\hat\omega)=1/\mu(i\hat\omega)$
if it is given a small negative slope for $\hat\omega<\omega_0$. We still assume dilute media,
 $|\xi|\ll 1$, as above. 

Consider now the case of finite temperatures. Since the thermal Green
function is periodic in imaginary time, we can replace the Fourier
transform in the $T=0$ theory with Fourier series in imaginary
time. The transition to finite temperature theory is accomplished by
means of a discretization of the frequencies,
\begin{equation} 
\hat\omega\to\hat\omega_n=2\pi n/\beta, \qquad
x\to x_n=\hat\omega_n a,
\end{equation}
\label{22}
where $n$ is an integer and $\beta=1/T$ (we put $k_{\mathrm B}
=1$). The rule for going from frequency integral to sum over Matsubara
frequencies is
\begin{equation} 
\int_0^\infty dx \to t \sideset{}{'}\sum_{n=0}^\infty \quad ,\qquad t=2\pi a /\beta .
\end{equation}
\label{23}
Here $t$ is a nondimensional temperature, and the prime on the summation sign in
(23) means that the $n=0$ term is counted with half weight.

Since $x_n$ can be written as $nt$, we can express the Casimir free energy
 at finite temperatures as
\begin{equation} 
F(T)= - \frac{\xi^2t}{\pi a}
\sum_{l=1}^\infty\nu\sideset{}{'}\sum_{n=0}^\infty \lambda^2_l(nt)
\end{equation}
\label{24}
to $\mathcal {O} (\xi^2)$. The corresponding zero temperature expression,
still to be called $E$, is given by the first term to the right in Eq. (12).

The following delicate point ought to be noted. We know that the mean energy per mode
 of the electromagnetic field is $\varphi(\omega, \beta)=\frac{1}{2}\omega \coth (\frac{1}{2}\beta \omega)$. The function $\varphi(\omega, \beta)$ is to replace $\frac{1}{2}\omega$, when we construct the general contour integral for the Casimir energy $E(T)$ at finite temperatures. It is the function $\coth(\frac{1}{2}\beta \omega)$ which is responsible for the occurrence of the Matsubara frequencies along the imaginary frequency axis. Now, our formalism implies morover a transition from the energy $E(T)$ to the free energy $F(T)$. This is accomplished by means of a partial integration, making use of the relations
\[ \frac{\partial \varphi(\omega,\beta)}{\partial \omega}=
\frac {\partial}{\partial \beta}\left( \frac{\beta}{\omega}\varphi(\omega,\beta) \right),
~~~~E(T)=\frac{\partial}{\partial \beta}\left( \beta F(T)\right). \]
For this reason we have changed the energy symbol from {E} to $F$ in Eq. (24). At $T=0$, obviously $F(T=0)=E(T=0) \equiv E$. We thank I. Klich (personal communication) for helpful comments regarding this point.

Let us in this context also recall how the surface force density $f$ is calculated. Generally, as mentioned in section 2,
$f$ can be found as the difference between the
radial Maxwell stress tensor components on the outside and inside of
the surface $r=a$. The stress components are in turn constructed from
the two-point functions like $ \langle E_i({\bf r}) E_k({\bf
  r'})\rangle $ when ${\bf r}$ and ${\bf r'}$ are close together, at a
given temperature $T$. The appropriate energy function is found by
integrating $4\pi a^2 f$ over $a$, from initial position $a= \infty$,
to the final position where the radius of the sphere terminates at the
fixed value $a$. This process is to take place at constant
temperature. The appropriate energy function is accordingly the {\it
  free energy} $F(T)$. Thus we see that, whereas in force considerations at finite temperatures one is lead naturally to the free energy, while starting from energy considerations one has instead to make use of thermodynamical considerations to calculate  $F(T)$.

Mathematically, the case $n=0$ requires special attention. For this
purpose we observe the approximate expression for $s_l$ and $e_l$ at
low $x$:
\begin{equation} 
s_l(x) = \frac{\sqrt\pi}{\Gamma(\nu+1)}
\left(\frac x 2 \right)^{\left(\nu + \half\right)},\qquad
e_l(x) = \frac{\Gamma(\nu)}{\sqrt\pi}
\left(\frac x 2 \right)^{\left(-\nu + \half\right)}.
\end{equation}
\label{25}

From this it follows that $s_l(x)e_l(x)= x/2\nu$,  $\lambda_l(x)=
1/2\nu$, implying that the contribution to (24) from $n=0$ is
\begin{equation} 
F(T, n=0)=-\frac{\xi^2 t}{8\pi a}\sum_{l=1}^\infty\frac 1\nu.
\end{equation}
\label{26}

We add the contribution from $n\ge 1$, restricting ourselves to the
dominant term in the uniform asymptotic expansion of $\lambda_l$
\cite{Brevik98}:
\begin{equation}  
\lambda_l(x)= \frac{(1+x^2/\nu^2)^{-3/2}}{2\nu}
\left[1 + {\cal O} \left( \frac 1{\nu^2} \right) \right] .
\end{equation}
\label{27}
Thereby
\begin{equation} 
F(T)= -\frac{\xi^2t}{8\pi a}
\sum_{l=1}^{l_0} \frac 1\nu 
\left[ 1 + \sum_{n=1}^\infty \frac{2\nu^6}{(\nu^2+ n^2t^2)^3}\right].
\end{equation}
\label{28}

We have here truncated the sum over $l$ at $l=l_0$ in order to avoid
divergences. It is easy to give an order-of-magnitude estimate of
$l_0$: Our dispersion relation (21) implies that photons having
frequencies $\omega$ higher than $\omega_0$ do not ``see'' the sphere
at all. A photon of limiting frequency $\omega_0$ just touching the
surface of the sphere has an angular momentum equal to $\omega_0 a=x_0$.
This type of argument has repeatedly been used in Casimir calculations
\cite{Candelas82,Brevik90,Brevik94}. One might put $l_0=c x_0$
where $c$ is a coefficient of order unity. However, in view if
simplicity, and since we are able to give only an estimate of the
order of magnitude anyhow, we put henceforth $c=1$. As upper
limit in the summation we thus take
\begin{equation}  l_0=x_0. \end{equation}
\label{29}

Now taking into account the exact summation formula derived in Eq.(A3)
in \cite{Brevik89} ( with $p$ an arbitrary constant):
\begin{equation} 
  \begin{split}
    \sum_{n=1}^\infty \frac 1{(n^2+p^2)^3} = 
    &-\frac 1{2p^6} + \frac{3\pi}{16p^5}\coth\pi p \\
    &+ \frac{3\pi^2}{16p^4}\frac 1{\sinh^2\pi p} +
    \frac{\pi^3}{8p^3}\frac{\coth\pi p}{\sinh^2 \pi p},
  \end{split}
\end{equation}
\label{30}
we can write Eq. (28) as
\begin{equation}
F(T)=
-\frac{3\xi^2}{64a}\sum_{l=1}^{x_0}
\left[ \coth\frac{\pi\nu}t 
+ \frac {\pi\nu/t}{\sinh^2\pi\nu/t} +
\frac 2
3\left(\frac{\pi\nu}{t}\right)^2\frac{\coth\pi\nu/t}{\sinh^2\pi\nu/t}
\right];
\end{equation}
\label{31}
the first term on the right hand side of (30) drops out.

This expression can be calculated, with $x_0$ as an input parameter,
to show how $F(T)$ varies with $t$. It is of interest first
to examine the case of low temperatures. Let us assume that 
\begin{equation} 
t\ll 1 ,
\end{equation}
\label{32}
implying that $\pi\nu/t \gg 1$. Then the first term in (31)
dominates,  and can be replaced by unity. We obtain approximately
\begin{equation}
F(t\to 0) = -\frac{3\xi^2}{64a}x_0
.\end{equation}
\label{33}

This negative expression establishes a constant low-temperature
plateau for the Casimir free energy. 

It is of interest to compare this
result with that obtained from the slightly more complicated though
physically more realistic Lorentz (or Sellmeir) dispersion relation.
The latter, when taken only along the imaginary frequency axis, can be
written as
\begin{equation} 
\mu(i\hat\omega)-1=\frac{\chi_0}{1+\hat\omega^2/\omega_0^2}=\frac{\chi_0}{1+x^2/x_0^2}   
,\end{equation}
\label{34}
where $\chi_0$ is  the zero frequency magnetic
susceptibility. (The damping term is omitted.) 
Here $\hat\omega=\omega_0$ represents a ``soft'' cutoff.  
This case was considered in \cite{Brevik88}, for the zero-temperature case.
For a dilute medium, $|\chi_0| \ll 1$, the result of the calculation in
\cite{Brevik88} can be written  
\begin{equation} 
F(t\to 0) \rightarrow E= -\frac{\xi^2}{16a}\left( x_0-\frac 3 4 \right)  
.\end{equation}
\label{35}
This formula assumes mathematically, strictly speaking, that $x_0\gg
1$ in addition to $t\ll 1$. The first condition is however not very
stringent; it turns out numerically that (35) holds to better than 1
percent accuracy even when $x_0$ is as low as about 4. We thus see
that the dominant first term in Eq. (35) agrees qualitatively well
with Eq. (33), the latter obtained on basis of the ``square''
dispersion relation (21). The latter expression is $3/4$ of the
Lorentz-based expression $E \simeq -(\xi^2/16a)x_0$.  This is a
satisfactory result, in view of the semi-quantitative agreement
between the meaning of the parameters $x_0$ occuring in (16) and in
(34).

The second term in Eq. (35), $E(\rm{nondisp})=(3/64)\xi^2$, is the
characteristic positive Casimir energy (corresponding to a repulsive
force component) for a {\it nondispersive} dilute ball satisfying the
condition $\epsilon \mu =1$, at zero temperature.  Cf., for instance,
\cite{Milton96}.

The other limiting case of interest is that of high temperatures. Let
us assume that
\begin{equation}
t\gg  x_0
,\end{equation}
\label{36}
implying that $\pi\nu/t\ll 1$ for all values of $l$ in
Eq. (31). Making use of $\sinh z = z$ and $\cosh z = 1$ for
small $z$, we obtain as leading term
\begin{equation}
F(t\gg  x_0) = 
-\frac {\xi^2t}{8\pi a}\sum_{l=1}^{x_0}\frac 1 \nu =-\frac{\xi^2}{4\beta}\sum_{l=1}^\infty \frac{1}{\nu}
.\end{equation}
\label{37}
This expression is independent of $a$. 
The high - temperature Casimir free energy, still negative, thus decreases
proportionally with $t$. This is qualitatively in agreement with \cite{Revzen97}
 and \cite{Feinberg99}.

\begin{figure}[h]
   \begin{center} 
     \includegraphics[scale = 0.5]{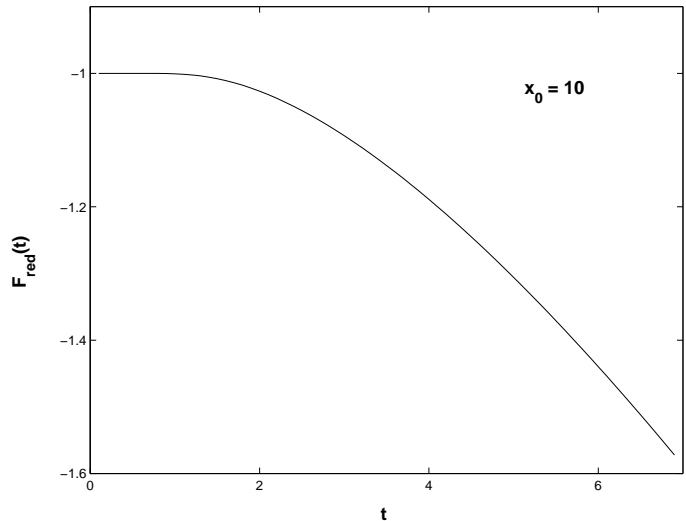} 
\end{center}

     \caption{ Using the "square" dispersion relation
     (21): Nondimensional free energy $F_{red}(t)$ as defined by
     equation (38), versus nondimensional temperature $t=2\pi a/\beta$
     when $x_0=10$.}   

\end{figure}

\begin{figure}[h]
   \begin{center}
   \includegraphics[scale = 0.5]{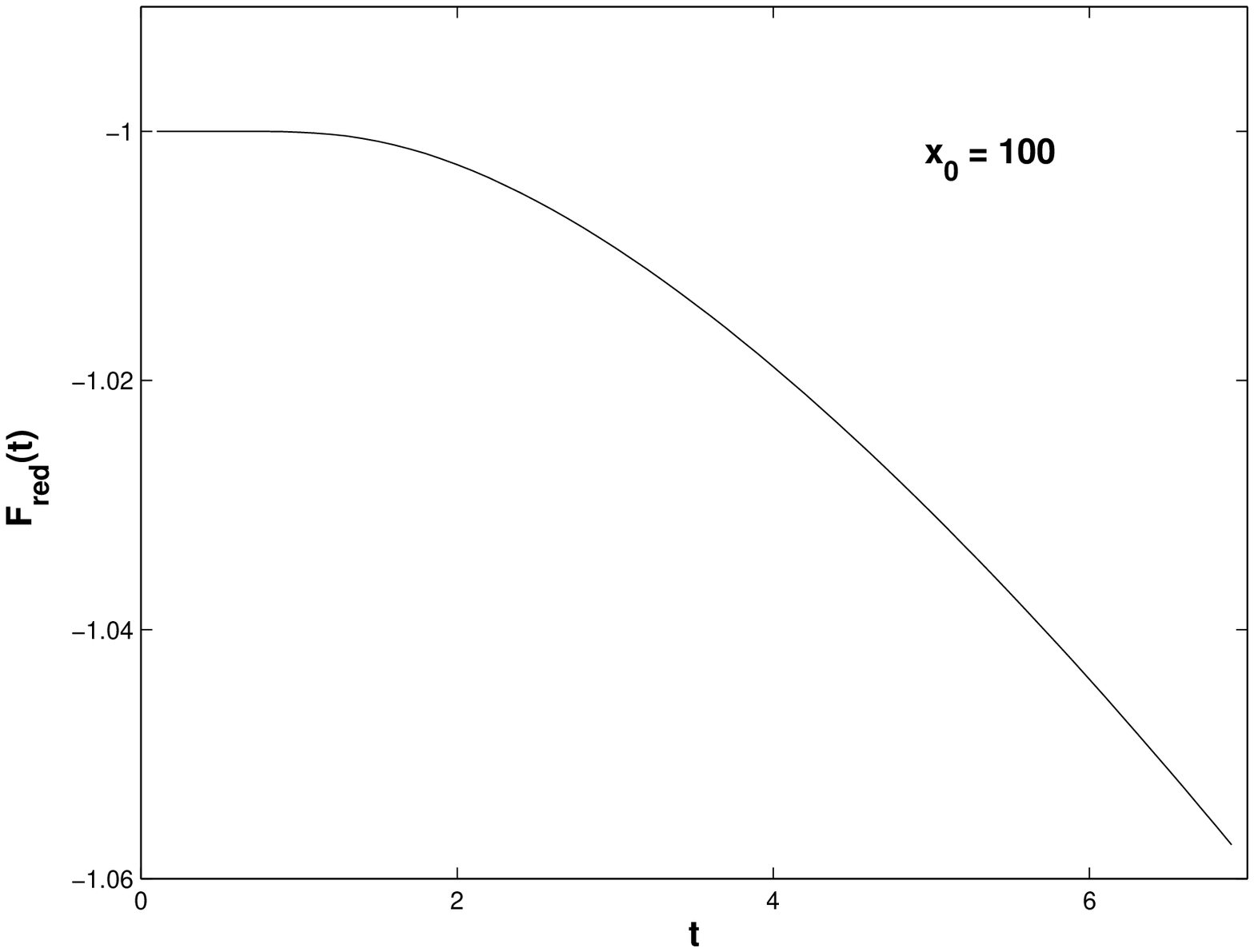}
   \end{center}
   \caption{  Same as figure 1, but with $x_0=100$.}
   
\end{figure}

\begin{figure}[h]
   \begin{center}
   \includegraphics[scale = 0.5]{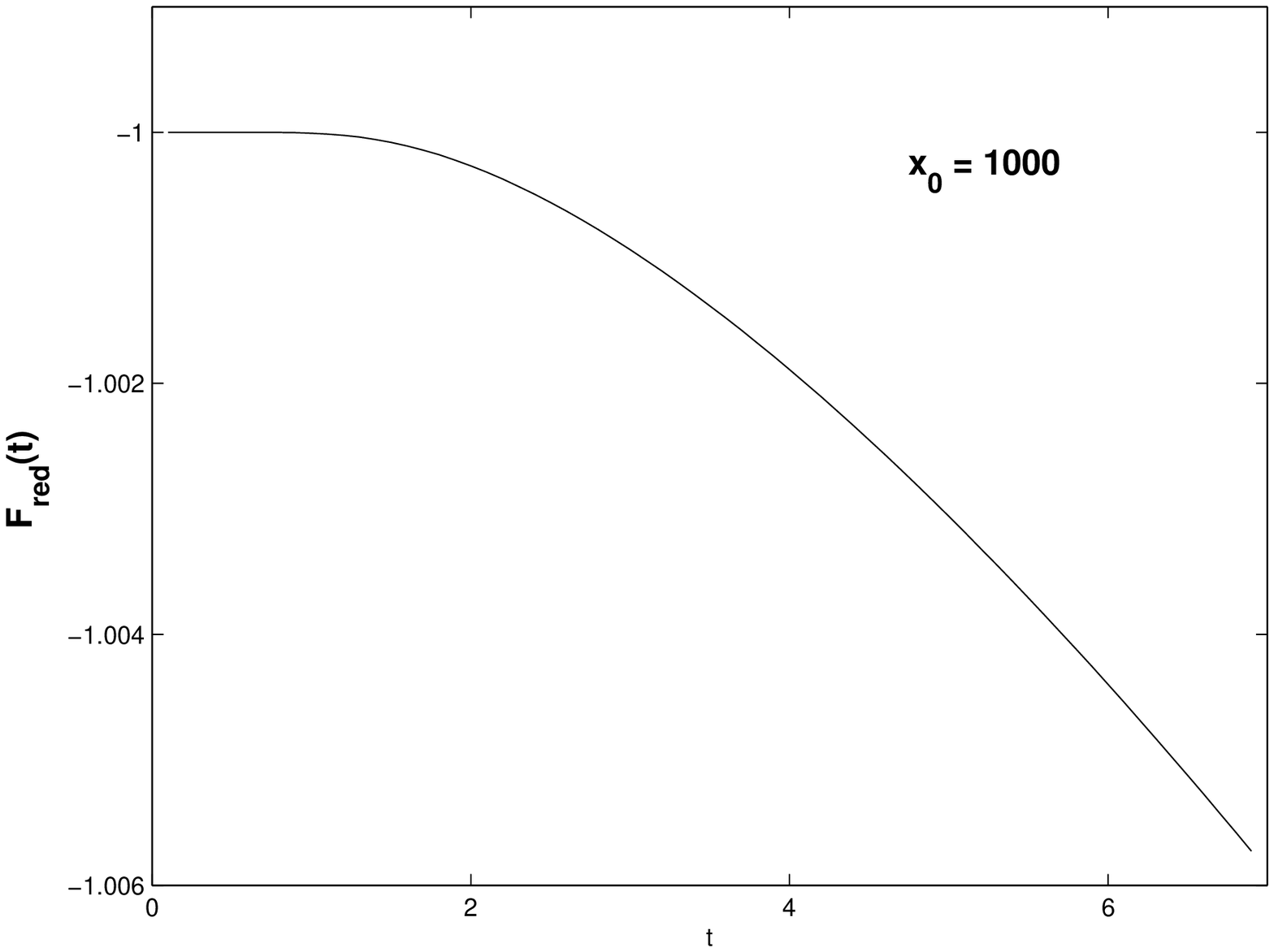}
   \end{center}
   \caption{  Same as figure 1, but with $x_0=1000$.}

\end{figure}

Figures 1 - 3  show how the nondimensional reduced Casimir free energy $F_{red}(t)$, 
defined by 
\begin{equation}
F_{red}(t) \equiv \frac{F(T)}{3\xi^2x_0/(64a)},
\end{equation}
\label{38}
varies with $t$, in the cases when $x_0=\{10,~100,~1000\}$. The low-temperature plateaus
evidently lie at $F_{red}(0)=-1$. These figures are given on a linear scale, thus exhibiting the 
linear behaviour predicted by Eq. (37) at high temperatures. It is of interest to check the accuracy
of (37) in cases when the condition (36) is reasonably well, but not extremely well, satisfied. Let us choose $x_0=10,~t=100$ as an example. From the expression
\begin{equation}
F_{red}(t\gg x_0)=-\frac{8t}{3\pi x_0}\sum_{l=1}^{x_0}\frac{1}{\nu}
\end{equation}
\label{39}
we obtain, taking into account that $\sum_1^{10}1/{\nu}=2.361749$,
that $F_{red}(x_0=10, t=100)=-20.047149.$ The full series solution
yields the number -20.047154.  The approximate formula (39) is thus in
this case extremely accurate.  Choosing $t=10$, we get from Eq. (39)
that $F_{red}(x_0=10, t=10)=-2.005$, whereas the series (31) yields
-2.091.  Even in this case, where the condition (36) obviously is
broken, the error in the formula (39) is thus only about 4 per
cent. Generally speaking, at high temperatures when the diluteness of
the medium becomes more pronounced, the Casimir free energy becomes
small, as we would expect.

\begin{figure}[h]
   \begin{center} 
   \includegraphics[scale = 0.5]{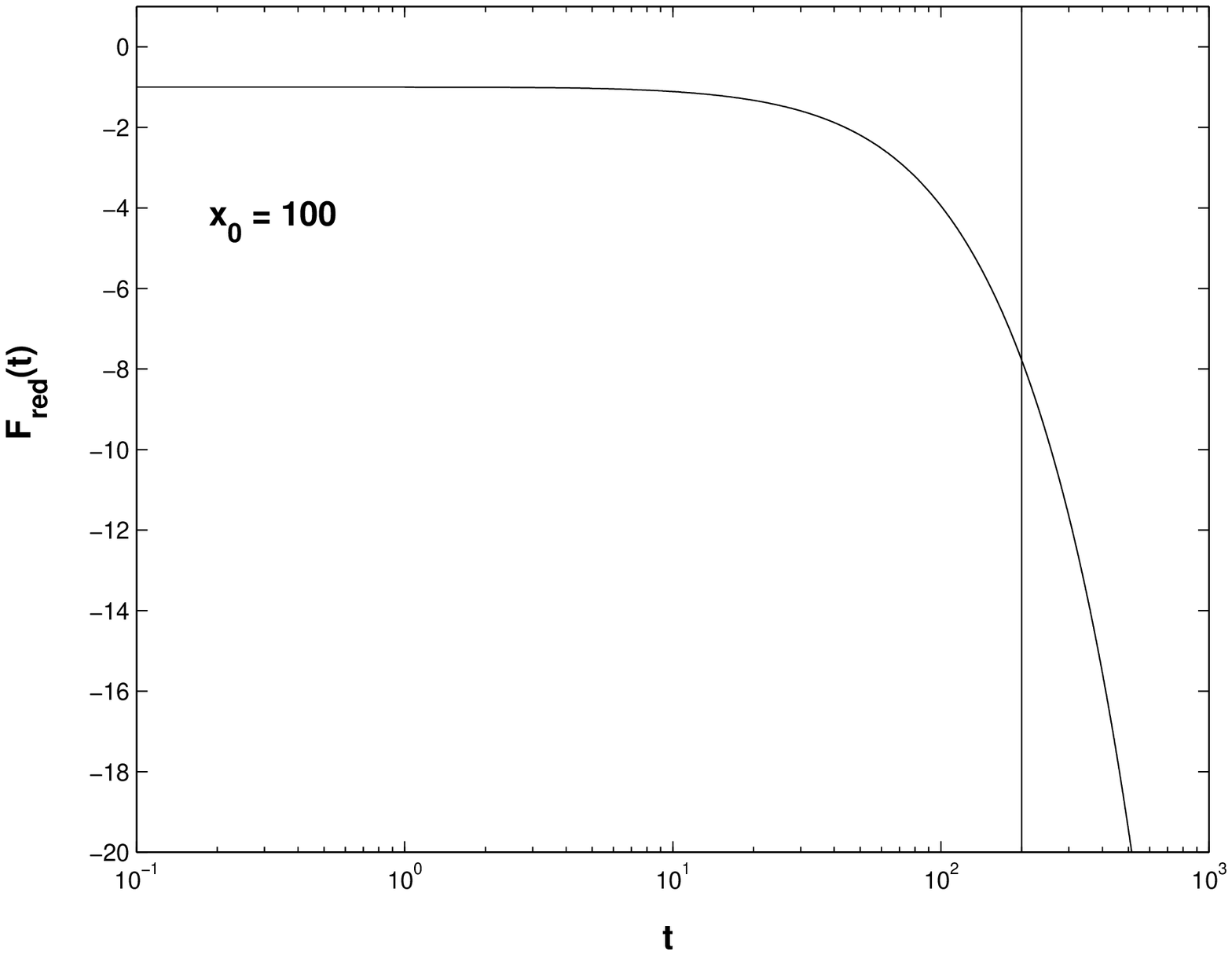}   
   \end{center}
   \caption{ Same as figure 2, but with a logarithmic temperature
    scale. Vertical line represents $t=2x_0$.}
\end{figure}

In connection with the temperature dependence of the free energy the
following physical argument should be noticed. From the dispersion
relations (21) or (34) we would expect that the influence from the
dispersion effect becomes important when the temperature becomes so
high that the most significant frequencies in the thermal radiation
field are of the same order as $\omega_0$. And as most significant
thermal frequency, it is natural to take the value $\omega=\omega_m$
corresponding to the {\it maximum} of the blackbody distribution. From
Wien's dispacement law we know that $\omega_m=2.8/\beta$. Thus, we
expect that dispersion becomes significant when $\omega \sim \omega_0
\sim 2.8/\beta$ which, in view of the definition $t=2\pi a/\beta$,
means that $t\sim 2x_0$.  Now, in Figs. 1-3 the linear scale chosen
for the temperature implies that the intervals covered on the ordinate
axes are small, and the mentioned effect becomes hidden (the
"shoulder" in each diagram appears to lie at a lower value of
$t$). The effect becomes however clear if we replace the linear
temperature scale by a logarithmic scale. Figure 4 shows, as an
example, such a diagram for the case $x_0=100$. Then the ordinate
interval becomes considerably larger, and we see that the prediction
$t \sim 2x_0$ for the position of the "shoulder turns out to be
reasonable.  (In the diagram we have drawn the line $t=2x_0$ to
emphasize the transitional region.)  The same behaviour was previously
found to hold in \cite{Brevik89}.

We note in passing that the high temperature entropy $S$ turns out to
be constant: from Eq. (37) we get
\begin{equation}
S(t \gg x_0)= -\frac{\partial F(T)}{\partial T}=\frac{\xi^2}{4}\sum_{l=1}^{x_0} \frac{1}{\nu}.
\end{equation}
\label{40}

\section{An improved dispersion relation}

Instead of the formal divergence encountered in the previous section
and the necessity to truncate the sum over $l$ in Eq. (28) at an upper
limit $l=l_0$, it is of interest to consider a mathematically more
involved but physically more correct dispersion relation, that leads
to a finite result for the free energy even after summing over all $l$
up to infinity.

Let us consider the following dispersion relation, being essentially of the Lorentz form but augmented by a term in the denominator that describes spatial dispersion: 
\begin{equation} 
\mu(i\hat\omega,l)-1=\frac{\chi_0}{1 + (x^2+l^2)/x_0^2},  
\end{equation}
\label{41}
$\chi_0$ being a constant. 
As in the previous section we assume
that the ball of radius $a$ is dilute ($|\chi_0 |\ll 1$), and that the
exterior region is a vacuum.  The spatial dispersion term $l^2/x_0^2$ ensures convergence when summing over all $l$. Since $\xi =(1-\mu)/(1+\mu)$ according to (\ref{eq:7}), we obtain to first order
in $\chi_0$, at finite temperatures,
\begin{equation} 
\xi\to\xi_{ln}=\frac{-\chi_0/2}{1 + (n^2t^2+l^2)/x_0^2}.   
\end{equation}
\label{42}
Instead of the expression (28) we now get 
\begin{equation}
F( T)=
-\frac{\xi_0^2x_0^4t}{8\pi a}
\sum_{l=1}^\infty
\left[ \frac 1 {\nu (x_0^2+l^2)^2} +
\sum_{n=1}^\infty
\frac{2\nu^5}{(x_0^2+l^2+n^2t^2)^2(\nu^2+n^2t^2)^3}
\right],
\end{equation}
\label{43}
where we have introduced the symbol $\xi_0=\chi_0/2$. We resolve  the last term into partial fractions,
and introduce the symbol $S(p;k)$:
\begin{equation}
S(p;k)=\sum_{n=1}^\infty (n^2+p^2)^{ -k} 
\end{equation}
\label{44}
Then, if $p$ and $q$ are arbitrary constants, we get
\begin{eqnarray}
\sum_{n=1} ^\infty \frac{1}{(n^2+p^2)^2(n^2+q^2)^3}&=& -\frac{3}{(p^2-q^2)^4}S(p;1)-\frac{1}{(p^2-q^2)^3}S(p;2)
\nonumber \\                            
                       &+& \frac{3}{(p^2-q^2)^4}S(q;1)-\frac{2}{(p^2-q^2)^3}S(q;2) \nonumber \\
                                                   &+&\frac{1}{(p^2-q^2)^2}S(q;3).
\end{eqnarray}
\label{45}    
From Appendix A in \cite{Brevik89} we write down the expressions for $
S(p;k) $; $ k=1,2$:
\begin{equation}
S(p;1)=-\frac{1}{2p^2}+\frac{\pi}{2p}\coth \pi p,
\end{equation}
\label{46}
\begin{equation}
S(p;2)=-\frac{1}{2p^4}+\frac{\pi}{4p^3}\coth \pi p +\frac{\pi^2}{4p^2}\frac{1}{\sinh ^2 \pi p}
\end{equation}
\label{47}
($S(p;3)$ was given in Eq. (30)). We define, in analogy with Eq. (38), the nondimensional reduced free energy to be
\begin{equation}
F_{red}(t) \equiv \frac{F(T)}{3\xi_0^2x_0/(64 a)},
\end{equation}
\label{48}
and obtain after some manipulations the following expression:
\begin{equation}
\begin{split}
F_{red}(t) & =   -\frac{16 x_0 ^3 t}{3\pi }\sum_{l=1}^\infty 
             \left\{
             \frac{1}{2\nu (x_0^2+l^2)^2} \right. \\
     & -   \frac{3\nu^5}{t^2 (x_0^2-l-\frac{1}{4})^4}
               \left[ S ( \frac{\sqrt{x_0^2+l^2}}{t};1 ) \,-\;S (
             \frac{\nu}{t};1) \right]  \\
     & -  \frac{\nu^5}{t^4 (x_0^2-l-\frac{1}{4})^3}\left[ S ( \frac{\sqrt{x_0^2+l^2}}{t};2 )
          + 2S(\frac{\nu}{t};2 )\right] \\
     & +  \left. \frac{\nu^5}{t^6 (x_0^2-l-\frac{1}{4})^2}S( \frac{\nu}{t};3)
            \right\}.
\end{split}
\end{equation}
\label{49} 
This expression is calculated numerically for two of the cases above, viz. $x_0=\{10,~100\}$. The results are shown in Fig. 5.

\begin{figure}[h]
   \begin{center} 
   \includegraphics[scale = 0.5]{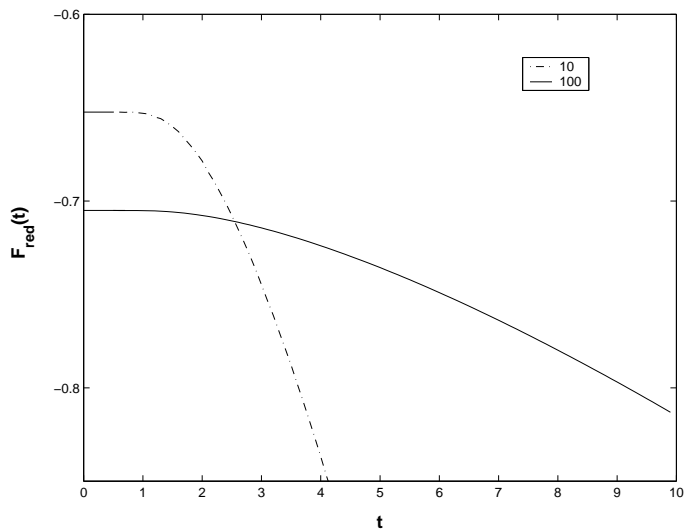}
   \end{center}

     \caption{ Using the spatial-dispersion relation
     (41): Nondimensional free energy $F_{red}(t)$ as defined by
     Eq. (48), versus nondimensional temperature $t=2\pi a/\beta$ when
     $x_0=\{10, ~100\}$. The low-temperature plateau is in accordance
     with Eq. (54).  }

\end{figure}

It is seen that the curves have roughly the same form as in Figs. 1 -
2. There is a definite low - temperature plateau, and for high values
of $t$ the curves tend towards a linear form. In any case, whatever we
adopt a square dispersion relation or a spatial - dispersion relation,
the sensitivity with respect to increasing values of $t$ is most
pronounced when $x_0$ is smallest. The limiting values for $t
\rightarrow 0$ in Fig. 5 are in agreement with those of Figs. 1 and 2
to within about 70 per cent. This is the order of magnitude -
agreement that we might expect, since the parameter $\xi_0$ made use
of in Eq. (48) is similar to, but not identical with, the
nondispersive quantity $\xi$ made use of in Eq. (38).

It is of interest to give analytic approximations for the case of low temperatures, $t\ll 1$. Then the quantities $p=\sqrt{x_0^2+l^2}/t$ and $q=\nu/t$ are large, and it becomes convenient to make use of the approximate formula \cite{Brevik89, Elizalde95}
\begin{equation}
\sum _{n=1}^\infty \frac{1}{(n^2+p^2)^k}=\frac{\sqrt{\pi}\Gamma(k-\frac{1}{2})}{2\Gamma(k)p^{2k-1}}
-\frac{1}{2p^{2k}}+\left( \frac{\pi}{p} \right)^k\frac{e^{-2\pi p}}{\Gamma(k)},
\end{equation}
\label{50}
from which it follows, in our notation, that
\begin{equation}
S(p;1)=\frac{\pi}{2p}-\frac{1}{2p^2}+\frac{\pi}{p}e^{-2\pi p},
\end{equation}
\label{51}
\begin{equation}
S(p;2)=\frac{\pi}{4p^3}-\frac{1}{2p^4}+\left( \frac {\pi}{p}\right)^2e^{-2\pi p},
\end{equation}
\label{52}
\begin{equation}
S(p;3)=\frac{3\pi}{16p^5}-
\frac{1}{2p^6}+\frac{1}{2} \left( \frac{\pi}{p} \right)^3 e^{-2\pi p}.
\end{equation}
\label{53}
Inserting Eqs. (51)-(53) into Eq. (49) we obtain a reasonable simple series expression for $F_{red}(t)$, at low temperatures. A further simplification can be obtained if we keep only the leading order terms in Eqs. (51)-(53). For $p \gg 1$ this means that we keep only the first term in each equation. We then get in this limit
\begin{eqnarray}
F_{red}(t \rightarrow 0)&=& x_0^3\sum_{l=1}^\infty \left\{\frac{8\nu^5}{(x_0^2-l-\frac{1}{4})^4}
\left( \frac{1}{\sqrt{x_0^2+l^2}}-\frac{1}{\nu} \right) \right. \nonumber \\
                        &+& \left. \frac{8\nu^5}{3(x_0^2-l-\frac{1}{4})^3}\left(
\frac{1}{2(x_0^2+l^2)^{3/2}}+\frac{1}{\nu^3} \right) 
                        -\frac{1}{(x_0^2-l-\frac{1}{4})^2} \right\}.
\end{eqnarray}
\label{54}  
In this expression, the temperature $t$ is no longer present. Numerical trials show that Eq. (54) is very accurate  for $t \ll 1$. Choosing $x_0=10,~t=1$, we obtain from Eq. (54) $F_{red}(x_0=10,~t=1)= -0.6526$, whereas the full formula (49) yields the number -0.6530, i. e., an error of 0.06 per cent. Even with $t=2$, where Eq. (49) yields the number -0.678, the error in Eq. (54) is increased to only about 3 per cent.

\section{Conclusion and final remarks}

Let us summarize as follows:

(1) For a relativistic medium, {\it i.e.}, a medium satisfying the
condition $\epsilon \mu =1$, the $T=0$ result for the Casimir energy
is most conveniently written in the form (8). This expression holds
for a nondispersive medium, $\delta=\hat{\tau}/a$ being the
time-splitting parameter. The permittivity is here arbitrary; the
medium need not be dilute. The regularization is made by subtracting
off the volume terms in the two scalar Green functions. This
regularization is natural, among else things because it permits one to
relate $E$ to the surface force density $f$ by the relation $\partial
E/\partial a=-4\pi a^2 f$. The equivalence between the Green function
approach and the mode sum approach is demonstrated explicitly.

(2) As an additional by-result at $T=0$, the expression (15)
generalizes the second order result for $E$ found in \cite{Brevik98}
up to the fourth order in the parameter $\xi$.

(3) Including the dispersive effect, the simplest dispersion relation
one can imagine is the "square" relation given in Eq. (21). The
finite-temperature expression for the free energy on the basis of this
dispersion relation is given in Eq. (31), with $x_0=\omega_0 a $ being
a high frequency cutoff which is in practice dependent on the detailed
structure of the medium. With reasonable accuracy $x_0$ can represent
the maximum $ l_0$ of the angular momentum variable $l$. Typical results
are shown in Figs. 1 - 4. In the limiting case $t \rightarrow 0$
they are in accordance with Eq. (33), and for $t \gg x_0$ they are in
accordance with Eq. (39). In particular, the free energy is for high
temperatures negative, and is a {\it linear} function of $t$. The
entropy is correspondingly a (positive) constant. These results are in
accordance with those of Refs. \cite{Revzen97} and \cite{Feinberg99}.
The transition region between low-temperature and high-temperature
theory can be determined with reasonable accuracy from a simple
physical argument.

These results are similar to those obtained on the basis of the
Lorentz dispersion relation in \cite{Brevik89}.

(4) One can take spatial dispersion into account, as we have done in
section 5. The basic dispersion relation is then Eq. (41). A
physically appealing feature of this kind of approach is that the sum
over angular momenta does not have to be truncated. Typical results
for this case are shown in Fig. 5.

(5) The condition $\epsilon\mu=1$ simplifies the theory, as noted
already in Ref. \cite{Brevik82}. For an ordinary dielectric ball
($\mu=1$) the calculation becomes more complex and difficult to
interpret \cite{Brevik99}.

(6) A final remark is called for, as regards the numerical computation
of the series in the case of spatial dispersion. The evaluation of the
expression (49) turned out to be more difficult than one might expect
beforehand. Thus, a simple use of Matlab turned out to be
insufficient. The problems seem to be associated with numerical noise,
caused by the "critical" terms for which $l \simeq x_0^2$. There are
relatively large individual terms that almost, but not quite,
compensate each other in the sum.  We managed to do the calculation
making use of double - precision MS-DOS Quick Basic. In practice,
adequate precision was found with inclusion of less than $x_0^2$
terms.

\bigskip
\bigskip
\bigskip

{\bf Acknowledgment}
\bigskip

We thank Michael Revzen and Israel Klich for valuable information
about the finite temperature problem.














\newpage
\begin{thebibliography}{99}

\bibitem{Plunien86}
Plunien, G., M\"{u}ller, B. and Greiner, W. 1986 {\it Phys. Reports}
{\bf 134}, 87.

\bibitem{Mostepanenko97} 
Mostepanenko, V. M. and Trunov, N. N. 1997 {\it The Casimir Effect
and its Applications} (London: Clarendon).

\bibitem{Milton99}
Milton, K. A. 1999 hep-th/9901011. Also in {\it Applied Field
Theory, Proc. 17th Symposium on Theoretical Physics}, eds. C. Lee,
H. Min and Q.-H. Park (Seoul: Chungburn, 1999).

\bibitem{Brevik82}
Brevik, I. and Kolbenstvedt, H. 1982  {\it Ann. Phys. NY} {\bf 143}, 179.

Brevik, I. and Kolbenstvedt, H. 1983  {\it Ann. Phys. NY} {\bf 149}, 237.

\bibitem{Brevik84}
Brevik, I. and Kolbenstvedt, H. 1984  {\it Can. J. Phys.} {\bf 62}, 805.

Brevik, I. and Kolbenstvedt, H. 1985  {\it Can. J. Phys.} {\bf 63}, 1409.

\bibitem{Milton78} 
Milton, K. A., DeRaad, L. L., Jr. and Schwinger, J.  1978 {\it Ann.
Phys. NY} {\bf 115}, 388.

\bibitem{Brevik88}
Brevik, I. and Einevoll, G. 1988 {\it Phys. Rev.} D {\bf 37}, 2977.

\bibitem{Brevik89}
Brevik, I. and Clausen, I. 1989  {\it Phys. Rev.} D {\bf 39}, 603.

\bibitem{Brevik87}
Brevik, I. 1987  {\it J. Phys. A: Math. Gen.} {\bf 20}, 5189.

\bibitem{Brevik90}
Brevik, I. and Sollie, R. 1990 {\it J. Math. Phys.} {\bf 31}, 1445.

\bibitem{Brevik94}
Brevik, I. and Nyland, G. H. 1994  {\it Ann. Phys. NY} {\bf 230}, 321.

\bibitem{Milton96}
Milton, K. A. 1996 {\it Proc. 3rd Workshop on Quantum Field Theory
Under the Influence of External Conditions} (Leipzig 1995), ed. M.
Bordag (Stuttgart: Teubner) p. 13.

\bibitem{Milton97}
Milton, K. A. and Ng, Y. J. 1997  {\it Phys. Rev.} E {\bf 55}, 4207.

\bibitem{Nesterenko97}
Nesterenko, V. V. and Pirozhenko, I. G. 1997 {\it Phys. Rev.} D {\bf
57}, 1284.

\bibitem{Brevik98} 
Brevik, I., Nesterenko, V. V. and Pirozhenko, I. G.  1998 {\it J.
Phys. A: Math. Gen.} {\bf 31}, 8661.

\bibitem{Lambiase98}
Lambiase, G., Nesterenko, V. V. and Bordag, M. 1999 {\it J. Math. Phys.} {\bf 40}, 6254.

Lambiase, G., Scarpetta, G. and Nesterenko, V. V. 1999 hep-th/9912176.


\bibitem{Nesterenko99}
Nesterenko, V. V. and Pirozhenko, I. G. 1999 {\it Phys. Rev. } D.
{\bf 60}, 125007.

\bibitem{Klich99}
Klich, I. 2000  {\it Phys. Rev. }D {\bf 61}, 025004.

\bibitem{Klich00}
Klich, I. 2000 {\it Casimir Energy for Spherical and Cylindrical Boundary Conditions}, MSc Thesis (Haifa: Israel Institute of Technology).

\bibitem{lifshitz55}
Lifshitz, E. M. 1955 {\it Zh. Eksp. Teor. Fiz.} {\bf 29}, 94 [{\it Sov. Phys. - JETP (USA)} {\bf 2}, 73 (1956)].

\bibitem{fierz60}
Fierz, M. 1960  {\it Helv. Phys. Acta} {\bf 33}, 855.

\bibitem{sauer62}
Sauer, F. 1962 Ph.D. thesis (University of G\"{o}ttingen).

\bibitem{mehra67}
Mehra, J. 1967  {\it Physica} {\bf 37}, 145.

\bibitem{Brown69}
Brown, L. S. and Maclay, G. J. 1969  {\it Phys. Rev.} {\bf 184}, 1272.

\bibitem{Milton78a}
Milton, K. A., DeRaad, L. L., Jr. and Schwinger, J. 1978 {\it Ann.
Phys. NY} {\bf 115}, 1.

\bibitem{Revzen97}
Revzen, M., Opher, R., Opher, M. and Mann, A. 1997 {\it J. Phys. A:
Math. Gen.} {\bf 30}, 7783.

\bibitem{Feinberg99}
Feinberg, J., Mann, A. and Revzen, M. 1999  hep-th/9908149.

\bibitem{Candelas82}
Candelas, P. 1982  {\it Ann. Phys. NY} {\bf 143}, 241.

\bibitem{Milton80}
Milton, K. A. 1980  {\it Ann. Phys. NY} {\bf 127}, 49.

\bibitem{brevik00}
Brevik, I., Jensen, B. and Milton, K. A. 2000  hep-th/0004041.

\bibitem{Lambiase96}
Lambiase, G. and Nesterenko, V. V. 1996 {\it Phys. Rev.} D {\bf 54},
6387.

\bibitem{Nesterenko97a}
Nesterenko, V. V. and Pirozhenko, I. G. 1997 {\it J. Math. Phys.}
{\bf 38}, 6265.

\bibitem{Landau84}
Landau, L. D. and Lifshitz, E. M. 1984 {\it Electrodynamics of
Continuous Media}, 2nd ed.  (Oxford: Pergamon) p. 280.

\bibitem{Elizalde95}
Elizalde, E. 1995  {\it Ten Physical Applications of Spectral Zeta Functions} (Berlin: Springer).

\bibitem{Brevik99}
Brevik, I. and Marachevsky, V. N. 1999  {\it Phys. Rev.} D{ \bf 60}, 085006.


\end{thebibliography}
\end{document}